\documentclass[10pt,conference]{IEEEtran}
\makeatletter
\def\ps@headings{%
\def\@oddhead{\mbox{}\scriptsize\rightmark \hfil \thepage}%
\def\@evenhead{\scriptsize\thepage \hfil \leftmark\mbox{}}%
\def\@oddfoot{}%
\def\@evenfoot{}}
\makeatother
\IEEEoverridecommandlockouts
\usepackage[left=0.63in,right=0.65in,top=0.75in,bottom=1.05in]{geometry}
\usepackage{amsfonts}
\usepackage{cite}
\usepackage{amsmath,amssymb,amsfonts}
\usepackage{algorithmic}
\usepackage{textcomp}
\usepackage{multirow}
\usepackage{graphicx}
\usepackage[table,xcdraw]{xcolor}
\usepackage{comment}
\usepackage{url} 
\newcommand{\SKC}{\text{SKC}}
\newcommand{\TA}{\text{TA}}
\newcommand{\TE}{\text{TE}}
\renewcommand{\S}{\text{\bf S}}
\newcommand{\G}{\text{\bf G}}
\newcommand{\U}{\text{\bf U}}
\renewcommand{\L}{\text{\bf L}}
\newcommand{\BS}{\text{\sf BS}}
\newcommand{\PL}{\text{\sf PL}}
\newcommand{\GS}{\text{\sf GS}}
\newcommand{\MC}{\text{\sf MC}}
\newcommand{\DPC}{\text{\sf DPC}}
\newcommand{\RT}{\text{\sf RT}}
\renewcommand{\SS}{\text{\bf SS}}
\newcommand{\GG}{\text{\bf GG}}
\newcommand{\SG}{\text{\bf SG}}
\newcommand{\SU}{\text{\bf SU}}
\newcommand{\GU}{\text{\bf GU}}
\newcommand{\UU}{\text{\bf UU}}
\newcommand{\C}{\text{\sc C}}
\newcommand{\I}{\text{\sc I}}
\newcommand{\A}{\text{\sc A}}

\usepackage{xcolor} 
\definecolor{LightGray}{gray}{0.9}
\newtheorem{insight}{Insight}
\newtheorem{definition}{Definition}
\definecolor{LightGray}{gray}{0.97}
\def\BibTeX{{\rm B\kern-.05em{\sc i\kern-.025em b}\kern-.08em
    T\kern-.1667em\lower.7ex\hbox{E}\kern-.125emX}}

\newcommand{\ignore}[1]{}

\makeatletter
\def\ps@IEEEtitlepagestyle{%
  \def\@oddfoot{\mycopyrightnotice}%
  \def\@oddhead{\hbox{}\@IEEEheaderstyle\leftmark\hfil\thepage}\relax
  \def\@evenhead{\@IEEEheaderstyle\thepage\hfil\leftmark\hbox{}}\relax
  \def\@evenfoot{}%
}
\def\mycopyrightnotice{%
  \begin{minipage}{\textwidth}
  \centering \scriptsize
  Copyright~\copyright~2023 IEEE. Personal use of this material is permitted. Permission from IEEE must be obtained for all other uses, in any current or future media, including reprinting/republishing this material for advertising or promotional purposes, creating new collective works, for resale or redistribution to servers or lists, or reuse of any copyrighted component of this work in other works.
  \end{minipage}
}
\makeatother

\begin{document}

\title{Characterizing Cyber Attacks against Space Systems with Missing Data: Framework and Case Study}

\author{
     \IEEEauthorblockN{Ekzhin Ear, Jose L. C. Remy, Antonia Feffer, and Shouhuai Xu}
     \IEEEauthorblockA{\em{Department of Computer Science} \\
     University of Colorado Colorado Springs \\
     \{eear, jcastano, afeffer2, sxu\}@uccs.edu
    }
}

\maketitle

\begin{abstract}
Cybersecurity of space systems is an emerging topic, but there is no single dataset that documents cyber attacks against space systems that have occurred in the past. These incidents are often scattered in media reports while missing many details, which we dub the {\em missing-data} problem. Nevertheless, even ``low-quality” datasets containing such reports would be extremely valuable because of the dearth of space cybersecurity data and the sensitivity of space systems which are often restricted from disclosure by governments. This prompts a research question: How can we characterize real-world cyber attacks against space systems? In this paper, we address the problem by proposing a framework, including metrics, while also addressing the missing-data problem, by ``extrapolating” the missing data in a principled fashion. To show the usefulness of the framework, we extract data for 72 cyber attacks against space systems and show how to extrapolate this ``low-quality” dataset to derive 4,076 attack technique kill chains. Our findings include: cyber attacks against space systems are getting increasingly sophisticated; and, successful protection against on-path and social engineering attacks
could have prevented 80\% of the attacks.

\end{abstract}

\begin{IEEEkeywords}
Space cybersecurity, satellite security incidents, cybersecurity metrics, cyber threat model, ATT\&CK, SPARTA
\end{IEEEkeywords}


\section{Introduction}\label{sec:intro}

Space systems, such as satellites, spacecraft, and space stations, are man-made objects that orbit the planet. 
They have become an underpinning of modern society because their missions include services that support many land, air, maritime, and cyber operations, such as Positioning, Navigation, and Timing (PNT) for the global stock market \cite{madry2015applications}, Satellite Communications (SATCOM) for global beyond-line-of-sight (BLOS) terrestrial voice and data requirements, and remote sensing for space domain awareness and planetary defense (e.g., detecting and deflecting large debris from hitting Earth). 

The space domain is interwoven with and enabled by the cyber domain, with real-time
cyber-physical systems comprising the space segment. Consequently, cyber attacks can affect
space systems, as evidenced by numerous space-related security incidents \cite{SpaceSecurityInfo, pavur2022building, fritz2013satellite, falco2021security}. 
Space incidents have occurred as early as 1977, with the hijacking of a satellite's audio transmission to broadcast the
attacker's own message \cite{fritz2013satellite}. In 1998, a U.S.-German RoSat (sensing satellite) experienced a malfunction that led to the satellite turning its x-ray sensor 
towards the sun, causing
permanent damage.
While it is debatable whether this incident was caused by cyber attacks,
the 
cyber attack
at the Goddard Space Flight Centre, where the RoSat is controlled, 
shows that cyber attacks can cause physical damage to space systems \cite{Wess2021}. 
Although cyber threats against space systems have become a reality, there is no systematic understanding of   cyber threats against space systems, likely due to the lack of data.

In this paper, we initiate the study on the problem of characterizing {\em real-world} cyber attacks against space systems {\em with missing data}. 
Ideally, we should have well documented attacks with significantly detailed descriptions (e.g., through digital forensics) to 
serve as input to this characterization study.
Unfortunately, cyber attacks, especially against space systems, are rarely well-documented owing to a variety of reasons (e.g., sensitivity), which explains why we must embrace the fact of missing data. The problem is important to deepen our understanding of cyber attacks against space systems, which are not yet understood, 
and to gain insights into making future
space systems secure.


\noindent{\bf Our contributions}. We make two technical contributions. 
First, we propose a novel framework to characterize real-world cyber attacks against space systems {\em with missing data}. The framework has three features: (i) It is {\em general} because it can accommodate both cyber attacks against space systems that occurred in the past and attacks that may occur in the future. 
(ii) It is {\em practical} because it offers approaches to deal with missing details of attacks, which is often the case with real-world datasets.
At a high level, our idea is to leverage the Aerospace Space Attack Research and Tactic Analysis (SPARTA) \cite{SPARTA} and the MITRE Adversarial Tactics, Techniques, and Common Knowledge (ATT\&CK) \cite{ATTCK} frameworks to extrapolate the missing details of attacks.
(iii) It offers two {\em metrics} for measuring {\em attack consequence} and {\em attack sophistication}, where each metric is a hierarchy of sub-metrics. These metrics may also be of independent value.

Second, we apply the framework to characterize the cyber attacks that are described in a dataset, which is prepared by this study and is, to our knowledge, the first comprehensive dataset of cyber attacks against space systems. The dataset is extracted from four publicly available datasets documenting space-related incidents,
which are not geared toward cyber attacks.
We manually extract
72 cyber attacks against space systems. 
Since many details of these 72 attacks are missing, we extrapolate them into 72 {\em attack tactic} chains (i.e., one attack tactic chain per attack) and 4,076 {\em attack technique} chains; we will make this dataset publicly available.
This allows us to draw a number of insights, such as: attacks against space systems can be effectively mitigated by hardening the ground segment; average-sophisticated cyber attacks can be effective against space systems, but attacks are getting increasingly sophisticated because of the increasing employment of defense; successful protection against on-path and social
engineering attacks could have prevented 80\% of the attacks.


\ignore{
(i) Our contribution at the means level: We propose an innovative framework to characterize cyber threats against space systems, which include the impact, complexity and capability perspective while leveraging the Space ATT\&CK framework. 
(ii) To facilitate the analysis, we ``artificially'' extrapolate the dataset  which can be leveraged to analyze other incidents than what are presented in the dataset we analyze.
(iii) We draw insights based on the analysis, such as ....:
(iv) We shed light on how to design solutions to securing space systems. 
}

\noindent{\bf Paper outline}.
Section \ref{sec:framework} describes the framework.
Section~\ref{sec:casestudy} presents a case study.
Section~\ref{sec:limits} discusses limitations. 
Section~\ref{sec:related_works} reviews prior studies. 
Section~\ref{sec:conclusion} concludes the paper.

\section{Framework}\label{sec:framework}





\subsection{Framework Requirements and Overview} 


To analyze cyber attacks against space systems, 
we need a framework
that satisfies the following requirements.
\begin{itemize}
\item {\bf{Requirement 0}:} The framework is {\em general} enough to accommodate all of the past cyber attacks (i.e., the ones that have been observed) and possible future attacks against space systems.

\item {\bf Requirement 1}: The framework is {\em practical} by accommodating real-world datasets with missing attack details.

\item {\bf{Requirement 2}:} The framework provides metrics for characterizing cyber attacks against space systems.
\end{itemize}
To address Requirement 0, we propose a system model that can be leveraged to describe past and future attacks. To address Requirement 1, we propose methods to ``extrapolate'' a given dataset to incorporate {\em hypothetical but plausible} missing details.  
To address Requirement 2, we define {\em attack consequence} and {\em attack sophistication} metrics.

\vspace{-1em}
\begin{figure}[!htbp]
\centering
\includegraphics[width=.4\textwidth]{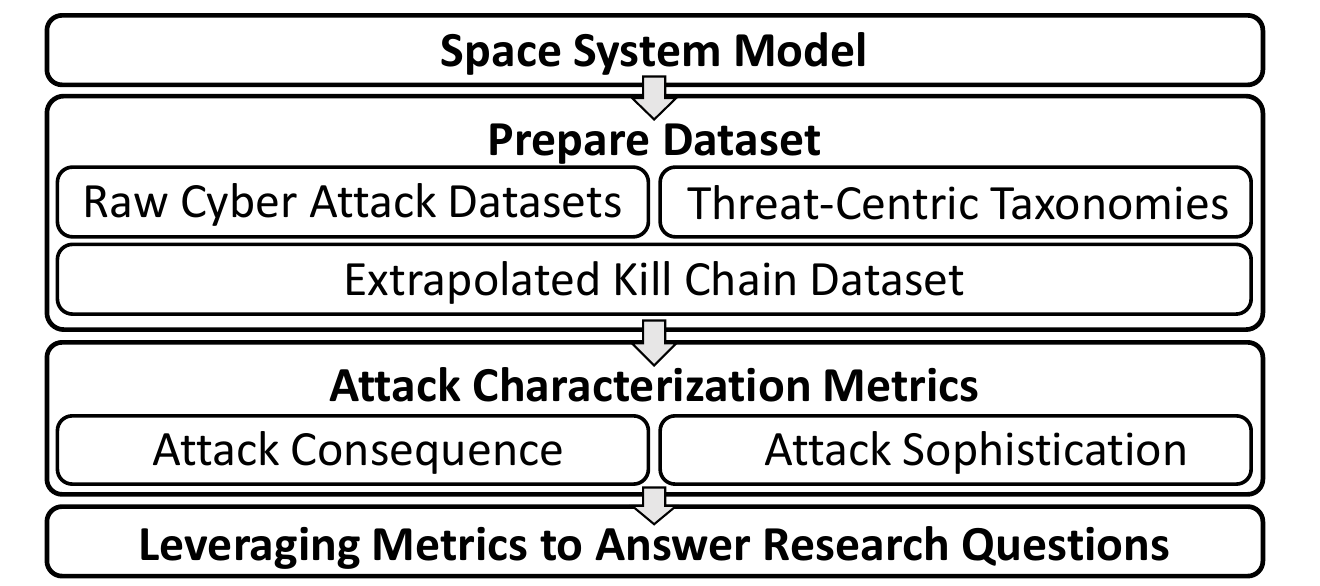}
\vspace{-1em}
\caption{The framework.}
\label{fig:framework}
\end{figure}

\vspace{-0.5em}
Fig.~\ref{fig:framework} highlights the framework, which has four major components: (i) designing the system model; (ii) preparing the dataset; 
(iii) defining metrics to quantitatively characterize cyber attacks; and (iv) leveraging the metrics to answer research questions.

\subsection{System Model}

\begin{figure}[!htbp]
\centering{\includegraphics[width=\columnwidth]{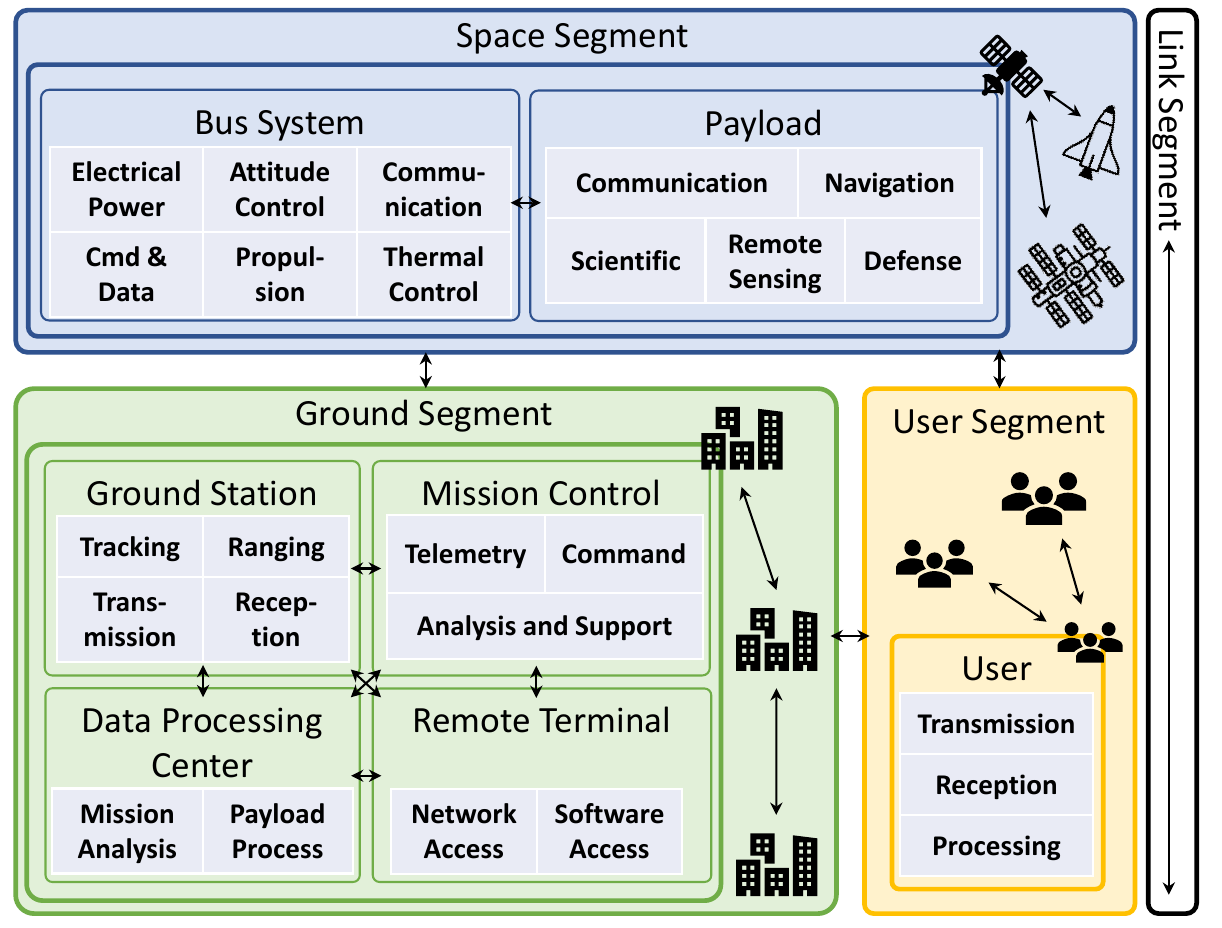}}
\vspace{-2em}
\caption{System model showing the four segments that comprise space systems.}
\label{fig:system_model}
\end{figure}

We observe that space systems often consist of four segments: {\em space}, {\em link}, {\em ground}, and {\em user} \cite{tedeschi2022satellite, guo2021survey, fritz2013satellite}. This prompts us to propose the system model described in Fig.~\ref{fig:system_model}. 

In the space segment, there are multitudes of space systems (e.g., satellites, spacecraft, and space stations). 
A space system {\em Bus System} facilitates  tracking, telemetry, and command requirements (TT\&C) and typically contains the following components \cite{chippalkatti2021recent, nguyen2020future}: {\em electrical power}, {\em attitude control}, {\em communication}, {\em command and data handling}, {\em propulsion}, and {\em thermal control}.
The {\em Payload} of space systems include communication (e.g., antennas and transmitters for relaying voice and data), navigation (e.g., Global Navigation Satellite System (GNSS) receivers 
for position and timing), scientific experiment (e.g., telescopes and spectrometers for research), remote sensing (e.g., sensors and cameras for terrestrial environmental monitoring), and national security (e.g., military equipment).

\vspace{-1em}

\vspace{-0.5em}
In the ground segment, {\em Ground Stations} contain the hardware and software to transmit and receive RF communication signals, as well as tracking and ranging space systems. {\em Mission Control} centers process telemetry data to assess the health of space systems, send commands to control space systems, conduct analyses to plan orbital maneuvers and assess conjunctions, and manage other aspects of space system operations. {\em Data Processing Centers} conduct deeper analyses of space system missions and process payload data. {\em Remote Terminals} provide a light-weight software stack and network connectivity to other elements of the ground segment.

The user segment is 
geographically dispersed across continents and oceans, often requiring space system services around the clock.
GNSS is an example of such a service used across the land, air, and maritime domains in automobiles, airplanes, and ships across the world. The user segment typically receives data directly from GNSS satellites while satellite communications (SATCOM) users also transmit traffic to satellites. 

The link segment is concerned with inter- (e.g., satellite to ground) and intra-segment (satellite to satellite) data connections. Satellites can communicate by passing data through the link segment with other satellites, ground stations, as well as directly to user terminals, potentially via its payload (e.g., for voice and data transfer)  or bus (i.e., for TT\&C). It possesses a variety of physical and especially electromagnetic properties across the spectrum. For our purpose, it suffices to consider the link segment from a cyber perspective that is concerned with the data that transits the link segment, rather than the physical properties of the link segment. 

\subsection{Preparing the Dataset}

\ignore{

We identify an innovative perspective, which is the {\em cyber threat characteristics}. 
We define this through four attributes (and accompanying metrics): impact $\to$ complexity $\to$ capability strength. 
Given our system model and the available raw real-world data described in Section~\ref{sec:context}, we consider the impacts of real-world cyber threat actors against various elements and segments of space systems. These impacts imply a level of complexity that threat actors must overcome to achieve their objectives. Their ability to do this implies their capability strength, from which we can derive the minimal capability requirement for threat actors to achieve their desired impacts.

}



\subsubsection{Properties of Ideal Datasets}

We inherit the following terms from ATT\&CK \cite{strom2018mitre}:
{\it attack tactic} specifies {\em what} an attacker wants to achieve; {\it attack technique} specifies {\em how} an attacker achieves the objective of an attack tactic;
{\it attack procedure} is an instantiation of an attack technique. Inspired by ATT\&CK, we propose the following variant concepts: {\it space cyber kill chain} is a vector containing attack tactics, attack techniques, and attack procedures that may traverse segments of a space system to achieve attack objectives
against the space system; and, {\it space attack campaign} is an 
instantiation
of a space cyber kill chain to achieve certain attack objectives.
Note that attack procedure is to attack technique what space attack campaign is to space cyber kill chain.


It would be ideal that a dataset of cyber attacks against space systems possesses the following properties:
\begin{itemize}
\item {\em Comprehensiveness}: This property deals with the coverage of a dataset, namely the aspects that are important for describing cyber attacks against space systems. Ideally, a dataset should contain every phase of attack including attack tactics, techniques, and procedures.

\item {\em Accuracy}: The details about attack tactics, 
techniques, and 
procedures described in a dataset should be accurate, leaving no room for ambiguity or misinterpretation. This property is also hard to guarantee because descriptions in real-world datasets are often ambiguous.

\item {\em Zero Missing Data}: This property means that every aspect of an attack that should be described in a dataset is indeed described in the dataset. The accuracy property implies that these descriptions should be accurate. 
\end{itemize} 
Unfortunately, real-world datasets 
miss a lot of due data.

\subsubsection{Addressing the Problem of Missing Data}\label{sec:addressingMissingData}
We propose leveraging threat-centric cybersecurity taxonomies, such as SPARTA \cite{SPARTA} and 
ATT\&CK
\cite{ATTCK} to populate the missing data (e.g., attack tactics and attack techniques). 
To illustrate this idea, we describe one approach to take the best of both SPARTA and ATT\&CK, which are chosen because they collectively contain 
details of attacker tactics, techniques, and procedures against space systems.
Nevertheless, our framework can incorporate various other threat-centric taxonomies. 

Our use of SPARTA and ATT\&CK is novel because we use them to guide us in populating the missing attack tactics and attack techniques of cyber attacks against space systems in a principled fashion.
We use them together to create more holistic 
space cyber kill chains that can encompass all segments of space systems. This is important because ATT\&CK focuses on enterprise systems (e.g., IT networks in the ground segment) while the majority of SPARTA elements focus on the space segment.
This is non-trivial because SPARTA and ATT\&CK do not provide an explicit approach to compile attack tactics into feasible space cyber kill chains. 
For this purpose, we define the following concepts:
\begin{itemize}
    \item {\it Objective activity}: an activity conducted by an attacker to achieve one attack goal 
    from the cyber attack. This aligns to the SPARTA and ATT\&CK tactics of {\it Exfiltration} and {\it Impact} and their associated attack techniques. 
    \item {\it Milestone activity}: an activity that directly progresses a space cyber kill chain towards a particular objective activity (as defined above). We categorize the SPARTA and ATT\&CK tactics of {\it Initial Access, Execution, Privilege Escalation,} and {\it Command \& Control} and their associated attack techniques as milestone activities.
    \item {\it Enabling activity}: an activity that establishes or modifies a current state of a system environment to facilitate a milestone activity. We categorize the SPARTA and ATT\&CK tactics of {\it Resource Development, Persistence, Defense Evasion, Lateral Movement,} and {\it Credential Access} as enabling activities.
    \item {\it Information discovery activity}: an activity that provides necessary information to support an enabling or milestone activity. We categorize the SPARTA and ATT\&CK tactics of {\it Reconnaissance, Discovery,} and {\it Collection} as information discovery activities.
    \item {\it Milestone block}: a vector comprised of information discovery, enabling, and milestone activities that comprehensively describes the requisite activities to accomplish a milestone activity. Milestone blocks are the building blocks for composing space cyber kill chains.
\end{itemize}
At a high level, these concepts will be used to construct space cyber kill chains as follows: {\em objective activities} are constructed first to represent the sketch of {\em space cyber kill chains}; the sketch is enriched with lower-level {\em milestone activities}, and then further enriched with milestone blocks consisting of the lowest level {\em enabling activities} and {\em information discovery activities}. Our method has three steps.



\ignore{

In the space domain, extrapolation of data has precedents. The process of extrapolating incident data to create an artificial dataset may be analogous to the difficulties of conducting Space Domain Awareness (SDA) tracking operations. By the end of 2022, U.S. Space Command was tracking more than 47,000 objects in space. There are not enough sensors to measure every orbit at all  times. Hence, various techniques, including, multiple-hypothesis tracking (MHT) are used to predict the tracks of these objects in between direct measurements \cite{blackman2004multiple, oliver2022event}. This prediction and hypothesizing process provides a constant and realistic picture of the space domain in lieu of direct sensor data. This application of MHT inspires us to supplement incomplete reporting of the space-related cybersecurity incidents by hypothesizing attack tactics and techniques to fill the gaps in the raw data and produce feasible and realistic cyber kill chain data. 

With a dataset that is richer in technical details, we are able to glean more insights into the threat actors' impacts, complexity, and strength.

}

First, we extract all of the {\it objective activities}, in terms of attack tactics and attack techniques, of the cyber attacks from a given cyber attack dataset. If this is not possible for an attack because of missing data, we use our domain expertise to identify the most probable (i.e., hypothetical but most plausible) ones for the attack.
For example, if the dataset does not give an explicit description of any objective activity but does describe that an attacker laterally moves into a sensitive development network, we can identify the objective activity as ``exfiltrating proprietary secrets.''
Then, we arrange the resulting objective activities
in chronological order. 

Second, we identify all of the {\em milestone activities} for each objective activity 
and arrange them in chronological order. If there is any gap because of missing data, we use our domain expertise to populate hypothetical but plausible 
additional milestone activities accordingly. For example, if the cyber attack data discusses the attacker escalating to administrator privileges to establish a command \& control (C2) channel via a web shell to exfiltrate sensitive data, then we can assume the attacker compromised a public-facing web application as an initial access milestone activity. 

Third, we construct a {\em milestone block} for each milestone activity by identifying all of the associated enabling and information discovery activities. 
We compensate for missing data by reasoning the requirements for each milestone block. For example, if an attacker established a C2 channel to an internal server that endured for months, we can assume the attacker conducted a reconnaissance information discovery activity to identify the network routing required and a persistence enabling activity to facilitate C2 beaconing.

The resulting 
attack tactics and techniques constitute space cyber kill chains.
By applying this method to space cyber attack datasets, we obtain 
a space cyber 
kill chain
dataset.

\subsection{Defining Metrics to Characterizing Attacks}

We define two metrics to characterize cyber attacks against space systems: {\em attack consequence} and {\em attack sophistication}, with each being a multi-dimensional vector which can be aggregated into a single number if desired. 

\subsubsection{Attack Consequence Metric}    
    
Given that space systems have four segments and that attack consequences may be manifested at some or all of the four segments, we define a vector of vectors, denoted by $(\vec{s}_\S,\vec{g}_\G,\vec{u}_\U,\vec{l}_\L)$, to represent the attack consequences to the space segment (\S), ground segment (\G), user segment (\U), and link segment (\L), respectively. The 4 vectors are defined as follows.

\noindent{\bf Consequence to space segment ($\vec{s}_\S$)}.
We define the attack consequence to a Space Segment $\S$ as $\vec{s}_\S=(\vec{s}_\BS,\vec{s}_\PL)$, where vector $\vec{s}_\BS$ denotes the consequence to the Bus System (\BS) and vector $\vec{s}_\PL$ denotes the consequence to the Payload ($\PL$). 
\begin{itemize}
\item We define $\vec{s}_\BS=(s_{\BS,1},\ldots,s_{\BS,6})$ as a vector of attack consequences to the 6 components of the Bus System: electrical power ($s_{\BS,1}$), attitude control ($s_{\BS,2}$), communication ($s_{\BS,3}$), command \& data ($s_{\BS,4}$), propulsion ($s_{\BS,5}$), and thermal control ($s_{\BS,6}$). We define $s_{\BS,j}\in [0,1]$, $1\leq j \leq 6$, as the degree of the functionality (i.e., availability) 
of the corresponding component being degraded because of the attack in question, where $s_{\BS,j}=0$ (or 1) means the functionality is 0\% (or 100\%) degraded. 

\item We define $\vec{s}_\PL=(s_{\PL,1},\ldots,s_{\PL,5})$ as a vector of attack consequences to the 5 payload components: communication ($s_{\PL,1}$), navigation ($s_{\PL,2}$), scientific application ($s_{\PL,3}$), remote sensing ($s_{\PL,4}$), and national security ($s_{\PL,5}$), with $s_{\PL,j}\in [0,1]$ in the same fashion as $s_{\BS,j}$.

\end{itemize}
The preceding definition of $\vec{s}_\S$ has several salient features, which also apply to the subsequent metrics corresponding to the other segments.
~First, we differentiate ``defining what to measure'' from ``how to measure what we need to measure.'' 
The present study addresses the former. Note also that our definitions remain valid when considering the inter-dependencies between different segments. Concerning the latter, it would be ideal that measurements of these metrics are provided to analysts for purposes such as our present study; or alternatively, there is a community-wide agreement on the degrees of degradation mentioned above. 
In the absence of both, one can use their own domain expertise to estimate these metrics.
~Second, we make the number of components specific to the system model described in Figure \ref{fig:system_model} (e.g., 6 components in the Bus System) to make the definitions easier to follow. The definitions can trivially generalize to accommodate an arbitrary number of components. 
~Third, the ``fine-granularity'' of $\vec{s}_\BS$ and $\vec{s}_\PL$ makes them suitable to compare the consequences of multiple attacks and to make statements like ``Attack 1 is more powerful than Attack 2.''
~Fourth, if desired and permissible to lose nuances from the fine granularity, one can aggregate the vector metrics 
into a single number. For example,  $\vec{s}_\BS$ can be aggregated into $\bar{s}_\BS$ via some mathematical function $f$, namely  $\bar{s}_\BS=f(s_{\BS,1},\ldots,s_{\BS,6})$, where $f$ can be for instance the (weighted) algebraic average function
which makes cybersecurity sense because they deal with the same property (i.e., availability).
We can similarly aggregate $\bar{s}_\BS$ and $\bar{s}_\PL$ 
into a single number.

\ignore{
We may need to aggregate $v\vec{s}_P=(s_{P,1},\ldots,s_{P,5})$ into a single number, denoted by $\bar{s}_P$, as some function $f_{s_P}$, namely  $\bar{s}_P=f_{s_P}(s_{P,1},\ldots,s_{P,5})$. For example, one specific instance is the algebraic average, meaning $\bar{s}_P=\frac{1}{5} \sum_{k=1}^5 s_{P,k}$. One immediate extension is to consider the varying importance of the component. Suppose the weighted importance of the components are respectively $w_{P,1},\ldots,w_{P,5}$ where $0\leq w_{P,k} \leq 1$ and $\sum_{k=1}^5 w_{P,k}=1$. Then, weighted average of consequence on the Payload Applications incurred by an attack can be computed as $\bar{s}_P=\frac{1}{5} \sum_{k=1}^5 w_{P,k} \times s_{P,k}$.
}

\noindent{\bf Consequence to ground segment ($\vec{g}_\G$)}.
We define the attack consequence to a Ground Segment $\G$ as $\vec{g}_\G=(\vec{g}_{\GS},\vec{g}_{\MC},$ $\vec{g}_{\DPC},\vec{g}_{\RT})$, where $\vec{g}_{\GS}$ is the consequence to the Ground Station (\GS), $\vec{g}_{\MC}$ is the consequence to Mission Control (\MC), $\vec{g}_{\DPC}$ is the consequence to the Data Processing Center (\DPC), $\vec{g}_{\RT}$ is the consequence to the Remote Terminal (\RT). 
\begin{itemize}
\item We define $\vec{g}_{\GS}=(g_{\GS,1},\ldots,g_{\GS,4})$ as a vector of consequences to the 4 components of the Ground Station: tracking ($g_{\GS,1}$), ranging ($g_{\GS,2}$), transmission ($g_{\GS,3}$), and reception  ($g_{\GS,4}$), where $g_{\GS,j}\in [0,1]$, $1\leq j \leq 4$, and 0 (1) means 0\% (100\%) functionality degradation. 

\ignore{

This ``fine-grained'' definition of $g_S$ can be used, for example, to compare the consequences incurred by two attacks so that we can make statement like ``attack 1 is more powerful than attack 2.''

We may need to aggregate $\vec{g}_S=(g_{S,1},\ldots,g_{S,4})$ into a single number, denoted by $\bar{g}_S$, as some function $f_{g_S}$, namely  $\bar{g}_S=f_{g_S}(g_{S,1},\ldots,g_{S,4})$. For example, one specific instance is the algebraic average, meaning $\bar{g}_S=\frac{1}{4} \sum_{j=1}^4 g_{S,j}$. One immediate extension is to consider the varying importance of the component. Suppose the weighted importance of the components are respectively $w_{S,1},\ldots,w_{S,4}$ where $0\leq w_{S,j} \leq 1$ and $\sum_{j=1}^4 w_{S,j}=1$. Then, weighted average of consequence on the Ground Station incurred by an attack can be computed as $\bar{g}_S=\frac{1}{4} \sum_{j=1}^4 w_{S,j} \times g_{S,j}$.

}

\item We define $\vec{g}_{\MC}=(g_{MC,1},g_{MC,2},g_{\MC,3})$ as a vector of consequences to the 3 components of the Mission Control: telemetry processing ($g_{\MC,1}$), commanding ($g_{\MC,2}$), and analysis and support ($g_{\MC,3}$), where $g_{\MC,j}\in [0,1]$, $1\leq j \leq 3$, in the same fashion as $g_{\GS,j}$.


\ignore{

Similar to the definition of $g_S$, this method can be used to compare the consequences incurred by multiple attacks in terms of relative power.

We may need to aggregate $\vec{g}_M=(g_{M,1},\ldots,g_{M,3})$ into a single number, denoted by $\bar{g}_M$, as some function $f_{g_M}$, namely  $\bar{g}_M=f_{g_M}(g_{M,1},\ldots,g_{M,3})$. For example, one specific instance is the algebraic average, meaning $\bar{g}_M=\frac{1}{3} \sum_{k=1}^3 g_{M,k}$. One immediate extension is to consider the varying importance of the component. Suppose the weighted importance of the components are respectively $w_{M,1},\ldots,w_{M,3}$ where $0\leq w_{M,k} \leq 1$ and $\sum_{k=1}^3 w_{M,k}=1$. Then, weighted average of consequence on the Mission Control incurred by an attack can be computed as $\bar{g}_M=\frac{1}{3} \sum_{k=1}^3 w_{M,k} \times g_{M,k}$.

}

\item We define $\vec{g}_{\DPC}=(g_{\DPC,1},g_{\DPC,2})$ as a vector of consequence to the 2 components of the Data Processing Center: mission analysis ($g_{\DPC,1}$) and payload processing ($g_{\DPC,2}$), where $g_{\DPC,j}\in [0,1]$ 
as similiar to $g_{\GS,j}$.

\ignore{

namely the degree of functionality of the component being degraded, where $g_{D,l}=0$ means the component is not affected by the attack in question, and $g_{D,l}=1$ means the component is completely degraded to a nonoperational status by the attack. Similar to the definition of $g_S$, this method can be used to compare the consequences incurred by multiple attacks in terms of relative power.

We may need to aggregate $\vec{g}_D=(g_{D,1},g_{D,2})$ into a single number, denoted by $\bar{g}_D$, as some function $f_{g_D}$, namely  $\bar{g}_D=f_{g_D}(g_{D,1},g_{D,2})$. For example, one specific instance is the algebraic average, meaning $\bar{g}_D=\frac{1}{2} \sum_{l=1}^2 g_{D,l}$. One immediate extension is to consider the varying importance of the component. Suppose the weighted importance of the components are respectively $w_{D,1},w_{D,2}$ where $0\leq w_{D,l} \leq 1$ and $\sum_{l=1}^2 w_{D,l}=1$. Then, weighted average of consequence on the Data Processing Center incurred by an attack can be computed as $\bar{g}_D=\frac{1}{2} \sum_{l=1}^2 w_{D,l} \times g_{D,l}$.

}

\item We define $\vec{g}_{\RT}=(g_{\RT,1},g_{\RT,2})$ as a vector of consequence to the components of the Remote Terminal: network access ($g_{\RT,1}$) and software access ($g_{\RT,2}$), where $g_{\RT,j}\in [0,1]$
as with $g_{\GS,j}$,

\ignore{

For $1\leq l \leq 2$, we define $g_{R,m}\in [0,1]$, namely the degree of functionality of the component being degraded, where $g_{R,m}=0$ means the component is not affected by the attack in question, and $g_{R,m}=1$ means the component is completely degraded to a nonoperational status by the attack. Similar to the definition of $g_S$, this method can be used to compare the consequences incurred by multiple attacks in terms of relative power.

We may need to aggregate $\vec{g}_R=(g_{R,1},g_{R,2})$ into a single number, denoted by $\bar{g}_R$, as some function $f_{g_R}$, namely  $\bar{g}_R=f_{g_D}(g_{R,1},g_{R,2})$. For example, one specific instance is the algebraic average, meaning $\bar{g}_R=\frac{1}{2} \sum_{m=1}^2 g_{R,m}$. One immediate extension is to consider the varying importance of the component. Suppose the weighted importance of the components are respectively $w_{R,1},w_{R,2}$ where $0\leq w_{R,m} \leq 1$ and $\sum_{m=1}^2 w_{R,m}=1$. Then, weighted average of consequence on the Remote Terminal incurred by an attack can be computed as $\bar{g}_R=\frac{1}{2} \sum_{m=1}^2 w_{R,m} \times g_{R,m}$.

}

\end{itemize}
Note that $\vec{g}_\G$ has the same salient features as $\vec{s}_\S$.

\noindent{\bf Consequence to user segment ($\vec{u}_\U$)}.
We define the attack consequence to a User Segment $\U$ as $\vec{u}_\U=({u}_1,{u}_2,{u}_3)$, whose elements respectively measure the consequence to components of $\U$: transmission ($u_1$), reception  ($u_2$), and processing ($u_3$), with $u_{j}\in [0,1]$, $1\leq j \leq 3$, as with $s_{\BS,j}$. 
Note that $\vec{u}_\U$ has the same salient features as $\vec{s}_\S$ or specifically $\vec{s}_\BS$.


\noindent{\bf Consequence to link segment ($\vec{l}_\L$)}.
We define the attack consequence to a Link Segment as $\vec{l}_\L=(\{\vec{l}_{\S}\},\{\vec{l}_{\G}\},\{\vec{l}_{\SS}\}, \{\vec{l}_{\GG}\}, $ $\{\vec{l}_{\SG}\}, \{\vec{l}_{\SU}\}, \{\vec{l}_{\GU}\},\{\vec{l}_{\UU}\})$, where the elements respectively correspond to a set of links within a space system
affected by an attack, within a ground wide-area network (WAN),
between two space systems,
between two ground WANs,
between a Space Segment and a Ground Segment, between a Space Segment and a User Segment, between a Ground Segment and a User Segment, and between two users.
We further define: 
\begin{itemize}
\item $\vec{l}_\S=(l_{\S,\C},l_{\S,\I},l_{\S,\A})$ as the consequences to a link between the Bus System and the Payload, where $l_{\S,\C},l_{\S,\I},l_{\S,\A}\in [0,1]$ are respectively the consequence to the confidentiality, integrity, and availability assurance of the link, with 0 (1) meaning 0\% (100\%) degradation. 

\item $\vec{l}_{\G}=(\vec{l}_{\GS,\MC},\vec{l}_{\GS,\DPC}, \vec{l}_{\GS,\RT},\vec{l}_{\MC,\DPC},
\vec{l}_{\MC,\RT},\vec{l}_{\DPC,\RT})$ as the consequences to links of components in a Ground Segment, where  $\vec{l}_{\GS,\MC}=({l}_{\GS,\MC;\C},{l}_{\GS,\MC;\I},{l}_{\GS,\MC;\A})$ $\in[0,1]^3$ are respectively the consequence to the confidentiality, integrity, and availability of the link between a Ground Station (\GS) and a Mission Control (\MC), with 0 (1) meaning 0\% (100\%) degradation. 
The other elements of $\vec{l}_{\G}$ are defined in the same fashion.
 
\item $\vec{l}_{\SS}=(\vec{l}_{\SS,\C},\vec{l}_{\SS,\I},\vec{l}_{\SS,\A})$ as the consequences to a link between two space systems,
where $\vec{l}_{\SS,\C},\vec{l}_{\SS,\I},\vec{l}_{\SS,\A}\in[0,1]^3$ are respectively the consequence to the confidentiality, integrity, and availability assurance of the link, with 0 (1) meaning 0\% (100\%) degradation.

\item $\vec{l}_{\GG}=(\vec{l}_{\GG,\C},\vec{l}_{\GG,\I},\vec{l}_{\GG,\A})$ as the consequences to a link between two ground WANs,
where $\vec{l}_{\SS,\C},\vec{l}_{\SS,\I},\vec{l}_{\SS,\A}\in[0,1]^3$ are respectively the consequence to the confidentiality, integrity, and availability assurance of the link, with 0 (1) meaning 0\% (100\%) degradation.

\item $\vec{l}_{\SG}=(\vec{l}_{\SG,\C},\vec{l}_{\SG,\I},\vec{l}_{\SG,\A})$ as the consequences to a link between a Space Segment and a Ground Segment, where $\vec{l}_{\SG,\C},\vec{l}_{\SG,\I},\vec{l}_{\SG,\A}\in[0,1]^3$ are the consequence to the confidentiality, integrity, and availability assurance of the link, with 0 (1) meaning 0\% (100\%) degradation.

\item $\vec{l}_{\SU}=(\vec{l}_{\SU,\C},\vec{l}_{\SU,\I},\vec{l}_{\SU,\A})$ as the consequences to a link between a Space Segment and User Segment, where $\vec{l}_{\SU,\C},\vec{l}_{\SU,\I},\vec{l}_{\SU,\A}\in[0,1]^3$ are the consequence to the confidentiality, integrity, and availability assurance of the link, with 0 (1) meaning 0\% (100\%) degradation.

\item $\vec{l}_{\GU}=(\vec{l}_{\GU,\C},\vec{l}_{\GU,\I},\vec{l}_{\GU,\A})$ as the consequences to a link between a Ground Segment and User Segment, where $\vec{l}_{\SS,\C},\vec{l}_{\SS,\I},\vec{l}_{\SS,\A}\in[0,1]^3$ are the consequence to the confidentiality, integrity, and availability assurance of the link, with 0 (1) meaning 0\% (100\%) degradation.

\item $\vec{l}_{\UU})=(\vec{l}_{\UU,\C},\vec{l}_{\UU,\I},\vec{l}_{\UU,\A})$ as the consequences to a link between two users,
where $\vec{l}_{\UU,\C},\vec{l}_{\UU,\I},\vec{l}_{\UU,\A}\in[0,1]^3$ are the consequence to the confidentiality, integrity, and availability assurance of the link, with 0 (1) meaning 0\% (100\%) degradation.

\end{itemize}
Note that $\l_\L$ has the same salient features as $\vec{s}_\S$ except how metrics may be aggregated. For example, it would not make good cybersecurity sense to aggregate $(l_{\S,\C},l_{\S,\I},l_{\S,\A})$ via a (weighted) algebraic average because confidentiality, integrity, and availability metrics describe different properties.  One may suggest to aggregate them via $1-[(1-l_{\S,\C})(1-l_{\S,\I})(1-l_{\S,\A})]$, which is reminiscent of the aggregation of Common Vulnerability Scoring System (CVSS) scores  \cite{first2019common}. However, 
this does not appear sound as this aggregation function appears rooted in Probability Theory, which however does not apply here because the events, even if $(l_{\S,\C},l_{\S,\I},l_{\S,\A})$ can be interpreted as probabilities, are not independent (e.g., the three assurances may be degraded at will by an attacker, rather than degrading independently of each other). Accordingly, we define:


\begin{definition}[Attack Consequence]
\label{definition:attack-consequence}
We define the attack consequence of a cyber attack against space systems as 
$$\left(\cup_{\S} \{\vec{s}_\S\},~\cup_{\G} \{\vec{g}_\G\},~\cup_{\U} \{\vec{u}_\U\}, ~\cup_{\L}  \{\vec{l}_{\L} \}\right),$$
where the union is over all the Space Segments (${\S}$), Ground Segments (${\G}$), User Segments (${\U}$), and Link Segments (${\L}$) that are affected by the attack.
\end{definition}

\subsubsection{Attack Sophistication Metric}

We define this metric to characterize how sophisticated a cyber attack against a space system is, through the lens of attack tactics and attack techniques.
Suppose for each cyber attack we extrapolate it into a set of $n$ space cyber kill chains, denoted by 
$\SKC = \{\SKC_1,\ldots,\SKC_{n}\}$,
where $\SKC_i$ ($1\leq i \leq n$) denotes the $i$th space cyber kill chain that is possibly associated with the cyber attack in question (when there is missing data describing the attack).
Suppose $\SKC_i$ is associated with a set of $u$ attack {\em tactics}, denoted by $\TA_i=\{\TA_{i,1},\ldots,\TA_{i,u}\}$, and with a set of $v$ attack {\em techniques}, denoted by $\TE_i=\{\TE_{i,1},\ldots,\TE_{i,v}\}$,
where ${\text{TA}_{i,j}}$ and  ${\text{TE}_{i,j}}$ are respectively associated with a given sophistication score $\alpha_{\text{TA}_{i,j}}$ and $\alpha_{\text{TE}_{i,j}}$.
For $\SKC_i$, we define its tactic sophistication as the maximum element among the element of $\TA_i$, namely $\max(\TA_i)=\max(\{\alpha_{\text{TA}_{i,1}},\ldots,\alpha_{\text{TA}_{i,u}}\})$, 
and define its attack technique sophistication as $\max(\TE_i)=\max(\{\alpha_{\text{TE}_{i,1}},\ldots,\alpha_{\text{TE}_{i,v}}\})$, which respectively correspond to the most sophisticated attack tactic and attack technique used by the attacker. Now we are ready to define:

\begin{definition}[Attack Sophistication] 
\label{definition:attack-sophistication}
For an attack described by a set of hypothetical but plausible space cyber kill chains $\SKC = \{\SKC_1,\ldots,\SKC_{n}\}$, we define its {\em possible highest sophistication} as vector $(\alpha_{\TA_+},\alpha_{\TE_+})$, where $\alpha_{\TA_+}=\max(\{\max(\TA_1),\ldots,\max(\TA_n)\})$ which corresponds to the most sophisticated attack tactic that is used among the possible kill chains, and $\alpha_{\TE_+}=\max(\{\max(\TE_1),\ldots,\max(\TE_n)\})$ which corresponds to the most sophisticated attack technique that is used among the possible kill chains.

We define its {\em possible lowest sophistication} as vector $(\alpha_{\TA_-},\alpha_{\TE_-)}$, where $\alpha_{\TA_-}=\min(\{\max(\TA_1),\ldots,$ $\max(\TA_n)\})$ which corresponds to the least sophisticated attack tactic that is necessary for the attack to succeed, and $\alpha_{\TE_-}=\min(\{\max(\TE_1),\ldots,\max(\TE_n)\})$ which corresponds to the least sophisticated attack technique that is necessary for the attack to succeed.
\end{definition}

Note that Definition \ref{definition:attack-sophistication} can be easily extended to define the metric of, for example, {\em possible average sophistication}.

\subsection{Leveraging the Metrics to Answer Research Questions}

Given a dataset of cyber attacks against space systems, the metrics defined above can be leveraged to specify interesting research questions, such as:
(i) What is the attack consequence of each cyber attack against space systems that has occurred in practice? Is there any trend exhibited by the attacks in terms of their attack consequence?
(ii) What is the attack sophistication of each cyber attack against space systems that has occurred in practice? Is there any trend exhibited by the attacks in terms of their attack sophistication or attack entry points?

\section{Case Study}\label{sec:casestudy}

We apply the framework to characterize cyber attacks against space systems based on a specific dataset of real-world space cyber incidents.


\subsection{Preparing a Dataset}


\subsubsection{Raw Datasets}

In search of publicly available datasets, we queried ``satellite incidents'', ``space attack data'', ``space incidents data'' in Google Scholar (with no limiting parameters) which returned 45 papers; we reviewed the top 50 results of a Google Search of ``satellite, space, incidents, attack, dataset''; and we consulted with industry practitioners for viable datasets. Consequently, there are only four publicly available datasets that contain information about space incidents 
\cite{SpaceSecurityInfo, pavur2022building, fritz2013satellite, falco2021security}, which will be referred to as {\em raw datasets}. They span four decades (1977-2019): 70 incidents reported in 
\cite{SpaceSecurityInfo}; 112 in \cite{pavur2022building}; 59 in \cite{fritz2013satellite}; and 1,847 in \cite{falco2021security}.
However, most incidents described in these raw datasets 
are not about cyber attacks against space systems (e.g., incidents related to solar flare anomalies).
Thus, we use domain expertise to manually extract the incidents by identifying those related to cyber attacks.
This leads to a dataset of 72 cyber attack incidents, among which 65 are extracted from three sources
\cite{SpaceSecurityInfo,pavur2022building,fritz2013satellite} that substantially overlap with each other, and 7 are extracted from \cite{falco2021security}. We refer to this resulting dataset
as the {\em raw cyber attack dataset}.


The (cyber attack) incidents are reported using different descriptions. Three datasets \cite{SpaceSecurityInfo,pavur2022building,falco2021security} describe incidents in terms of: (i) year of incident occurrence; (ii) category of the attack / incident (e.g., jamming); (iii) attack target
(e.g., the USA-193 recon satellite); (iv) attacker identification
(e.g., nation-state); (v) narrative of facts concerning the incident, which are all brief; (vi) high-level attack objective (e.g., state-espionage);
and, (vii) source references such as news articles reporting the incident.
The fourth dataset \cite{fritz2013satellite} only provides (i), (ii), (v), and (vii). Even so, they all miss some data.

Consequently,
it is difficult to gauge the veracity of the 72 attacks described in the raw cyber attack dataset
as there is a lack of information across all segments related to the space systems involved in the attacks to gain ground-truth about what actually transpired in these incidents. For example, in the RoSat incident, a report claims causation
between the cyber intrusion of the Goddard Space Flight Centre and the satellite's subsequent destructive maneuver \cite{fritz2013satellite, Wess2021, Schneier2008cyberattacks, Epstein2008}. However, an interview with a scientist on the RoSat project refutes this causation \cite{McDowell2011}, which another source resonates \cite{soesanto2021terra}.

\subsubsection{Extrapolating Raw Data with Possible Missing Information}

To make the {\em raw cyber attack dataset} suitable for our research, we address this missing-data problem {\em manually}, which is possible because there are only 72 cyber attacks.
As described in the framework, we leverage SPARTA and ATT\&CK to populate the missing data 
and construct attack tactic chains and attack technique chains, as follows.



We begin by reviewing each incident's narrative to construct a chronological vector of events and actions from both the attacker and the defender, because the defense can inform us of the necessary actions of the attacker. We review each incident's sources referenced and conduct our own search for additional details concerning the incident and add the new findings to our chronological vector of events. At least 25\% of the incidents contained sources that provided contradicting details of the associated events. When this occurred, we favored the accounts that made the most cybersecurity sense and provided the most details as this aligns more with the objective of the present study. We also consider the other attributes of the incident in the dataset, e.g., a nation-state actor versus a private actor, when characterizing our vector of events.
Then, we correlate the chronological vector of events to
SPARTA and ATT\&CK tactics and techniques to build a preliminary attack tactic chain and attack technique chain.
For example, a source for the RoSat incident describes the use of phished
account credentials to gain access to the ground station; this attack tactic can be mapped to the Initial Access tactic within the milestone activities category with the corresponding Valid Accounts: Local (T1078.003) technique.
It is the case for all 72 of the incidents that this chain is incomplete.

We populate missing attack tactics
by selecting {\em objective activities} of attack tactics (defined in Section~\ref{sec:addressingMissingData})
according to the context provided in the raw cyber attack data, and by making plausible assumptions about the system based on our domain expertise (e.g., cybersecurity hygiene, defensive posture, and network architecture). Accordingly, we add plausible but missing 
{\em milestone activities} and {\em blocks}
to support the objective activities.
The resulting series of tactics constitutes an attack campaign.
We repeat this step to create multiple attack campaign scenarios by modifying our assumptions (e.g., services running with elevated privileges do not require the attacker to employ the privilege escalation tactic and hence decreases the number of {\em milestone activities} required). This means that each attack will be extrapolated into multiple hypothetical but possible attack campaigns. 
For example, one attack campaign may involve gaining initial access, privilege escalation, and lateral movement while a second attack campaign can progress straight from initial access to lateral movement because of the exploited service running as root. 
    
We populate missing attack techniques by leveraging the attack campaigns extrapolated from an attack as each campaign describes the attack from beginning to end. We leverage our domain expertise to first identify hypothetical cyber threat actors for the campaign and then 
hypothetical but plausible attack techniques for each attack tactic according to 
the attack procedures in the ATT\&CK matrix for the particular threat actors
\cite{pennington2019getting}. In this process, we further consider the affinity of details in the raw cyber attack data to particular ATT\&CK or SPARTA techniques. For example, given that data confirms RoSat attackers used FTP bounce as an Exploit for Defense Evasion (ATT\&CK technique T1211) \cite{Wess2021}, it is plausible to determine that the subsequent attack tactic and technique would be Reconnaissance and Active Scanning (T1595), respectively.

\subsubsection{Generating Extrapolated Kill Chain Dataset}
We construct hypothetical but plausible space cyber kill chains 
for each attack by leveraging the populated attack tactics and techniques according to the constructed attack campaign mentioned above.
For example, an attack campaign we re-constructed for the RoSat incident contains 14 ATT\&CK or SPARTA attack tactics, 9 of which come from the raw cyber attack data, each with one associated attack technique per the reported details. Of the 
5 extrapolated 
attack tactics, we associate each with different choices of extrapolated attack techniques, ranging from 2 to 6 extrapolated techniques. Consequently, we populate 1,296 hypothetical but plausible kill chains corresponding to all the combinations.

\vspace{-1em}
\begin{figure}[!htbp]
\centering
\includegraphics[width=\columnwidth]{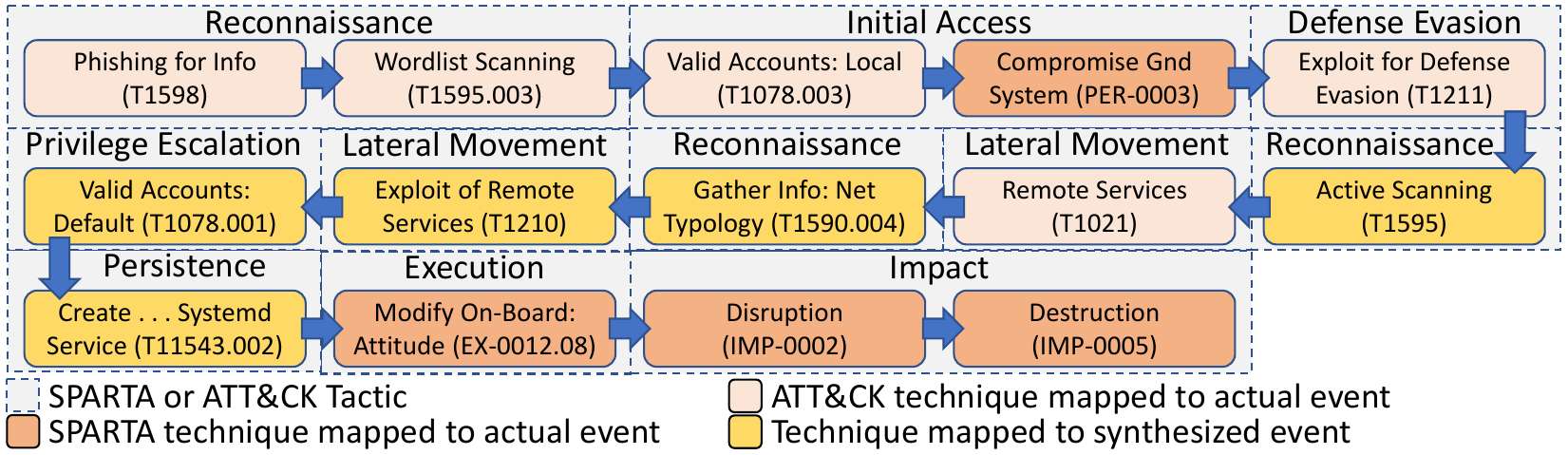}
\vspace{-2em}
\caption{One example kill chain (out of the 1,296 hypothetical but plausible chains) we constructed via our extrapolation for the RoSat 1998 attack.}
\label{fig:example-kill-chain}
\end{figure}

\vspace{-1em}
Fig.~\ref{fig:example-kill-chain} depicts one example of these kill chains constructed for the RoSat incident. If we follow the attack tactic chain, we see at a high level that the attackers conducted a thorough reconnaissance to gain initial access to a particular system at a ground station. They then conducted defense evasion and further reconnaissance to enable lateral movement deeper into the networks where they conducted further reconnaissance to laterally move into a position to affect the space segment. There they escalated their privileges to establish persistence on the network and enable execution of their cyber payloads to ultimately impact the RoSat satellite and destroy its x-ray sensing device. Following the attack technique chain would provide us more details on how these tactics could have been accomplished.



We organize the resulting hypothetical and plausible space cyber kill chains into what we call the
{\em Kill Chain Table} for each attack.
There are 4,076 space kill chains in total.
We will make this table publicly available.


\subsection{Characterizing Cyber Attacks}

We analyzed the 72 cyber attacks, divided into 8 categories:  25 jamming incidents, 17
Computer Network Exploitation (CNE) incidents, 16 hijacking incidents, 4 control	incidents, 3 Denial-of-Service (DoS) incidents, 3 eavesdropping incidents, 
1 theft-loss incident and 3 spoofing incidents. 
It is somewhat surprising to observe that
11 (15\%) attacks involved social engineering attacks, especially phishing. This means that cyber social engineering attacks are a threat to space cybersecurity.
We also observe that 71\% of the attacks fall into three related categories: political, state espionage, and criminal. 

\subsection{Leveraging the Metrics to Answer Research Questions}


\ignore{

{\color{red}What trends have been exhibited by cyber attacks against space systems in terms of attack impact, attack complexity, attack capabilities?}

...

Our resutls are limited in that our solution for RQ2 is specific to the particular dataset of this case study. Hence, our findings here may not be generalizable cross other datasets.

- high level insight: picture showing trend through the years
- number of attacks by type
{\color{purple}
The higher number of incidents that we have been able to compiled is 25 jamming incidents being followed by 18 CNE and 16 hijacking incidents. From that point on, the rest of categories ony contain very few incidents, varying from 4 incidents to 1 incident.

}
    - incorporate origin of attacks

}

\subsubsection{Attack Consequence Analysis}


We leverage Definition~\ref{definition:attack-consequence}, our space system model, our raw cyber attack dataset, and our extrapolated kill chain dataset to holistically consider the attack consequence of each cyber attack in our dataset, namely in terms of the cyber attack's consequence to the space, ground, user, and link segments. For each cyber attack, we apply our domain expertise to assess the attack consequence for each segment by assigning a score between 0 and 1 from least to most consequential. 

\begin{figure}[!htbp]
\centering
\includegraphics[width=.8\columnwidth]{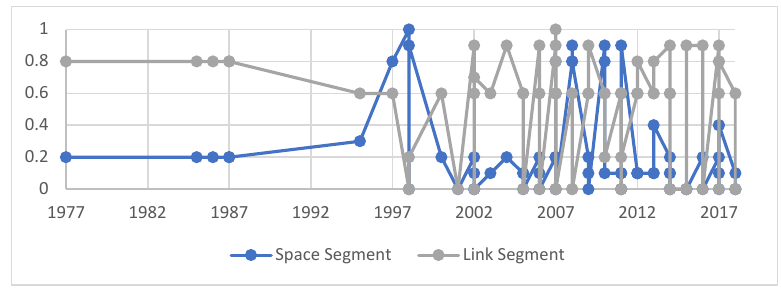}
\vspace{-1em}
\caption{Attack consequence in the space and link segments for the 72 attacks over time.}
\label{fig:consequence-space-link}
\end{figure}

Fig.~\ref{fig:consequence-space-link} illustrates that attack consequence for the space segment is generally low (i.e., 0.4 or below) with less than 15\% scoring 0.8 or higher: 5 jamming, 3 control, 1 theft, and 1 DOS attacks. The 1998 RoSat incident scored the highest as it demonstrated the ability to physically destroy an asset in the space segment via cyber means. Fig.~\ref{fig:consequence-space-link} also depicts a contrasting trend for the link segment. Two-thirds of the cyber attacks sustained an attack consequence score of 0.6 or greater. Of the most consequential attacks (0.9 or greater): 8 were hijacking and 2 were eavesdropping. The Russian Turla Hacking Group attained the highest score of 1.0 for their 2007 abuse of the link segment via asynchronous satellite internet connections to enable their global C2 reach.

\vspace{-1em}

\begin{figure}[!htbp]
\centering
\includegraphics[width=.8\columnwidth]{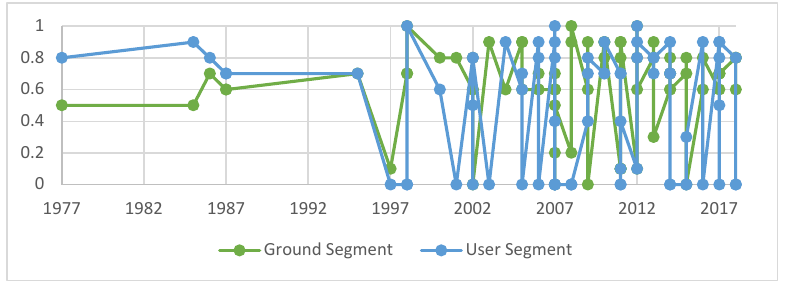}
\vspace{-1em}
\caption{Attack consequence in the ground and user segments for the 72 attacks over time.}
\label{fig:consequence-ground-user}
\end{figure}

\vspace{-0.5em}
Fig.~\ref{fig:consequence-ground-user} shows almost 80\% of the incidents have an attack consequence of 0.6 or greater for the ground segment while over 50\% of the incidents score 0.6 or greater against the user segment. These incidents cover all the categories of attacks, including those intended to compromise other segments, e.g., jamming and control attacks. This reveals how vital it is to protect the ground and user segments in order to protect the other segments in the space system model.

\vspace{-1em}

\vspace{-0.5em}

\noindent{\bf How can we prioritize the hardening of space system segments and components to mitigate the damage caused by cyber attacks?} 
We observe that 16 of 17 CNE and 3 of 4 control attacks possessed an attack consequence score of 0.6 or greater for the ground segment. Further, the extrapolated kill chain dataset reveals the victims' potentially weak security hygiene, e.g., default credentials, and unpatched public facing application vulnerabilities. This leads to the following insight:

\begin{insight}
CNE incidents can be mitigated by prioritizing hardening measures within the ground segment.
\end{insight}

\subsubsection{Attack Sophistication Analysis}


There are 14 attack tactics and 79 attack techniques in the extrapolated space kill chains corresponding to the 72 attacks.
For the 14 attack tactics, we manually score each from 0 to 1 considering that 0.5 is the average sophistication required to accomplish an attack tactic. For example, we score the {\em Initial Access} tactic as 0.5 because the majority of successful cyber attacks should gain initial access to its target. We assign a score of 0.8 or higher for the tactics requiring high sophistication. For example, the {\em Defense Evasion} tactic is scored at 0.9 as it requires additional effort and more advanced capabilities.
For the 79 attack techniques,
we assign the sophistication score of each while bearing in mind whether the score should be
less than, the same as, or higher than 
that of the associated attack tactic. 
Phishing is more commonplace and may require little technical capabilities when compared to other Initial Access techniques, and hence receives a score of 0.3.
We then compute attack sophistication according to Definition \ref{definition:attack-sophistication}.



\vspace{-1em}
\begin{figure}[!htbp]
\centering
\includegraphics[width=.9\columnwidth]{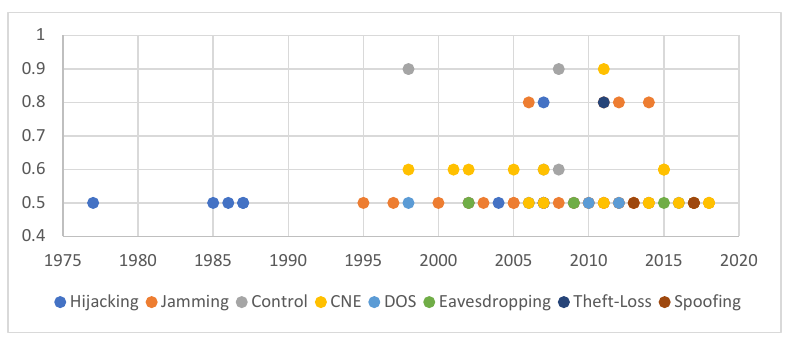}
\vspace{-1em}
\caption{Attack tactic sophistication of the 72 attacks over time. 
}
\label{fig:TA_soph_over_time}
\end{figure}

\vspace{-0.5em}
Fig.~\ref{fig:TA_soph_over_time} depicts the possible highest sophistication of the 72 attacks via $\alpha_{\TA_+}$, while noting that several plots overlap and hide other plots from view. We observe that hijacking attacks consistently scored 0.5, with one exception. In 2007,  the Russian Turla Hacking Group's employment of the C2 tactic (ATT\&CK ID TA0011) by leveraging SATCOM connections scored 0.8, which we consider as high sophistication. Jamming attacks also consistently scored 0.5 except for four attacks that attained a score of 0.8, each due to the successful employment of the Persistence tactic (SPARTA ID ST0005): Libyan cyber actors sustained jamming for six months in 2006 and again in 2011; protesters repeatedly jammed Thailand government television broadcasts in 2014; and North Korea successfully disrupted GPS signals, jamming a total of 553 aircraft over the course of a week. 50\% (2 of 4) of attacks to control a satellite attained high sophistication scores, in 1998 and 2008, namely for the successful employment of the Defense Evasion tactic (ATT\&CK ID TA0005) to overcome ground control station defenses. Almost half of the CNE attacks scored above 0.5 with one incident from 2011 that scored 0.9 by employing both Defense Evasion and Persistence tactics that enabled a series of 46 subsequent attacks against the ground segment. The only theft-loss incident in our dataset, also from 2011, scored 0.8 successfully employing the Persistence tactic for 2 years. All DOS, eavesdropping, and spoofing attacks in our dataset scored 0.5. Overall, attacks are getting more sophisticated.

\begin{figure}[!htbp]
\centering
\includegraphics[width=.9\columnwidth]{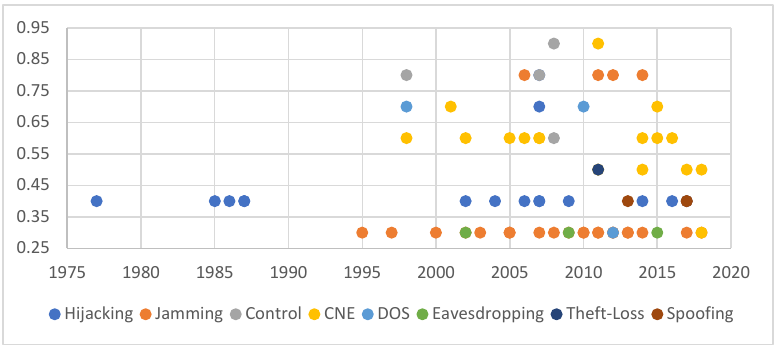}
\vspace{-1em}
\caption{Attack technique sophistication of the 72 attacks over time. 
}
\label{fig:TE_soph_over_time}
\end{figure}

Fig.~\ref{fig:TE_soph_over_time} depicts the possible highest sophistication of each attack via $\alpha_{\TE_+}$. We observe that $\alpha_{\TE_+}$ and $\alpha_{\TA_+}$ identify the same set of highly sophisticated cyber attacks. The notable attack techniques employed by these attacks to support the Defense Evasion tactic are: Indicator Removal (T1070) and Exploit (i.e., of a vulnerability) for Defense Evasion (T1211). The attacks that employed the Persistence tactic also employed the following attack techniques: Event Triggered Execution (T1546),  Create or Modify System Process (T1543), and Exploit Hardware/Firmware Corruption (EX-0005).

\vspace{-1em}


\vspace{-0.5em}
In comparing the results of our $\alpha_{\TA_+}$ and $\alpha_{\TE_+}$ measurements, we observe that they generally follow the same trend per incident,
but $\alpha_{\TE_+}$ scores are slightly more dispersed than $\alpha_{\TA_+}$. This is reasonable considering that attack tactics are at one level of abstraction higher than attack techniques. Hence, $\alpha_{\TE_+}$ will exhibit greater sensitivity in measurement. 
Nevertheless, 66\% of the incidents have both $\alpha_{\TA_+}$ and $\alpha_{\TE_+}$ scores of 0.5 or less. This leads to the following insight:
\begin{insight}
Cyber attacks of average sophistication can be successful against space systems.
\end{insight}



\noindent{\bf How many attacks would have been stopped by using simple countermeasures?}
We observe that 48 attacks employed techniques to establish an on-path attack position, and that the $\alpha_{\TE_+}$ score for these attacks are 0.4 or less. Though these attacks typically target the user segment, they leverage the weakness of the link segment. This leads to the insight:

\begin{insight}
Proper security of the link segment between the space and user segments (e.g., using cryptography) could have thwarted two-thirds of the observed cyber attacks. 
\end{insight}


We observe the sizeable frequency of social engineering techniques employed. The $\alpha_{\TE_+}$ score we assigned to the attacks that employed these techniques is 0.4. However, this is at the attack technique level of abstraction. We acknowledge that social engineering sophistication is highly nuanced at the implementation level, but simple defenses can be effective against phishing attacks, such as email filtering services, and application whitelisting. This leads to the following insight:

\begin{insight}
Traditional IT security controls against social engineering attacks could have thwarted 50\% (11/22) of the cyber attacks against
the ground segment.
Successful protection against {\em on-path} and {\em social engineering} attacks could have prevented 80\% of the cyber attacks against space systems.
\end{insight}

\section{Limitations}\label{sec:limits}

The present study has several limitations. First, we address the missing-data problem in a {\em manual}, rather than automated, fashion. 
Future research needs to investigate automated and objective methods for this process. 
Second, our metrics can be refined. For example, the attack consequence  metrics, except those associated with Link Segments, are geared toward  {\em availability} because the raw dataset lacks information about what kinds of confidential data are processed by these space systems. 
Third, we assume the measurements of the ``building-block'' metrics are given as input. While reasonable because of the focus of the present study, it is an important future research to investigate how to obtain these measurements, which may require a community effort. 
Fourth, the uncertainty associated with the populated attack tactics, attack techniques, space cyber kill chains, and space attack campaigns is not quantified. This issue is relevant because some scenarios may be more probable than others.
Fifth, our case study has limited generalizability due to the lack of publicly available data.

\section{Related Work}\label{sec:related_works}



There are studies on analyzing space-related incidents \cite{falco2021security,
pavur2022building,fritz2013satellite,boschetti2022space, pavur2020tale, soesanto2021terra}.
For example, \cite{pavur2022building} considers incidents in terms of the payload, signal, and ground aspects; \cite{fritz2013satellite} provides narrative descriptions concerning NASA, jamming, hijacking, and control attack categories; \cite{falco2021security} analyzes 1,847 space-related incidents according to their risk taxonomy for space. 
By contrast, we are the first to analyze cyber attacks against space systems, while preparing the first dataset with 72 cyber attacks.

There are studies on threat models for space applications, such as leveraging LEO constellations to jam GEO targets \cite{rawlins2022death}, creating attack trees against CubeSats \cite{falco2021cubesat}, demonstration of command injection via a software-defined radio \cite{lin2022defending}, characterizing the transmission layer's susceptibility to eavesdropping \cite{richardson2022ensuring}, leveraging the ATT\&CK framework \cite{ormrod2021cyber}, and nanosatellites as attack platforms \cite{pavur2021same}. By contrast, we analyze real-world cyber attacks, while aiming to leverage our findings to make future abstract threat models more realistic and holistic. Moreover, our framework is innovative, including metrics that have not appeared in the literature \cite{Pendleton16,XuSTRAM2018ACMCSUR,XuAgility2019,
XuSciSec2021SARR}.

\section{Conclusion}\label{sec:conclusion}
We presented an initial study on characterizing cyber attacks against space systems. 
We proposed an innovative framework with precisely defined metrics, while addressing the missing-data problem.
We prepared the first dataset of cyber attacks against space systems, including hypothetical but plausible attack details. By applying the framework to the dataset, we drew a number of insights. The limitations represent outstanding open problems for future research.   

\smallskip

\noindent{\bf Acknowledgement}. We thank the reviewers for their comments. This work was supported in part by Colorado State Bill 18-086. 

\vspace{-1em}


{\small

\bibliographystyle{ieeetr}
\bibliography{metrics}

}

\end{document}